\documentclass{aa}

\usepackage{enumitem}
\usepackage{graphicx}
\usepackage{txfonts}
\usepackage{natbib}
\bibpunct{(}{)}{;}{a}{}{,}

\usepackage{lscape}
\usepackage{longtable}
\usepackage{amssymb}

\def\G23{G023.01$-$00.41}

\usepackage{color}

\begin{document}

  \title{Discovery of a sub-Keplerian disk with jet around a 20\,M$_{\odot}$ young star}
  \subtitle{ALMA observations of G023.01$-$00.41}

  \author{A. Sanna\inst{1,2} \and A. K\"{o}lligan\inst{3} \and L. Moscadelli\inst{4} \and R. Kuiper\inst{3} 
              \and R. Cesaroni\inst{4} \and T. Pillai\inst{1} \and K.\,M. Menten\inst{1} \and Q. Zhang\inst{5} 
              \and A. Caratti\,o\,Garatti\inst{6} \and C. Goddi\inst{7} \and S. Leurini\inst{2} \and C. Carrasco-Gonz\'{a}lez\inst{8}}

                                      

   \institute{Max-Planck-Institut f\"{u}r Radioastronomie, Auf dem H\"{u}gel 69, 53121 Bonn, Germany \\
   \email{asanna@mpifr-bonn.mpg.de}
                 \and INAF, Osservatorio Astronomico di Cagliari, via della Scienza 5, 09047 Selargius (CA), Italy
                 \and Institute of Astronomy and Astrophysics, University of T\"{u}bingen, Auf der Morgenstelle 10, D-72076 T\"{u}bingen, Germany
                 \and INAF, Osservatorio Astrofisico di Arcetri, Largo E. Fermi 5, 50125 Firenze, Italy
                 \and Harvard-Smithsonian Center for Astrophysics, 60 Garden Street, Cambridge, MA 02138, USA
                 \and Dublin Institute for Advanced Studies, Astronomy \& Astrophysics Section, 31 Fitzwilliam Place, Dublin 2, Ireland
                 \and Department of Astrophysics/IMAPP, Radboud University Nijmegen, PO Box 9010, NL-6500 GL Nijmegen, the Netherlands
                 \and Instituto de Radioastronom\'{i}a y Astrof\'{i}sica UNAM, Apartado Postal 3-72 (Xangari), 58089 Morelia, Michoac\'{a}n, M\'{e}xico}

   \date{Received 11 May 2018; accepted 14 January 2019}


  \abstract{It is well established that Solar-mass stars gain mass via disk accretion, until the mass reservoir of the disk is exhausted and
    dispersed, or condenses into planetesimals. Accretion disks are intimately coupled with mass ejection via polar cavities, in the form of
    jets and less collimated winds, which allow mass accretion through the disk by removing a substantial fraction of its angular momentum. Whether
    disk accretion is the mechanism leading to the formation of stars with much higher masses is still unclear. Here, we are able to build a
    comprehensive picture for the formation of an O-type star, by directly imaging a molecular disk which rotates and undergoes infall around
    the central star, and drives a molecular jet which arises from the inner disk regions. The accretion disk is truncated between 2000--3000\,au,
    it has a mass of about a tenth of the central star mass, and is infalling towards the central star at a high rate ($6\times10^{-4}$\,M$_{\odot}$\,yr$^{-1}$),
    as to build up a very massive object. These findings, obtained with the Atacama Large Millimeter/submillimeter Array at 700\,au resolution,
    provide observational proof that young massive stars can form via disk accretion much like Solar-mass stars.}

   
   
   


   \maketitle
%

\section{Introduction}

Models of massive star formation in the disk accretion scenario predict that circumstellar disks could reach radii between 1000
and 2000\,au \citep[e.g.,][]{Krumholz2007,Kuiper2011,Harries2017}. Observationally, evidence for gas rotation near young stars with 
tens of Solar masses exists \citep[e.g.,][]{Zapata2010,Qiu2012,Johnston2015,Ilee2016,Beuther2017,Cesaroni2017}, although the amount of gas mass undergoing
rotation is a significant fraction of the star mass, and these envelopes, possibly hosting an inner disk, might be prone to fragmentation due
to self-gravity and/or develop spiral instabilities \citep[e.g.,][]{Kratter2010,Kuiper2011,Klassen2016,Chen2016,Meyer2017,Meyer2018}. 
The interplay between disks and jets provides a mechanism to ensure mass accretion through the disk. Notwithstanding their connection,
there is poor evidence of disk-jet systems in the inner few 1000\,au of stars with tens of Solar masses \citep[e.g.,][]{Beltran2016},
such as those resolved around Solar- and intermediate-mass stars \citep[e.g.,][]{Lee2017b,Lee2017a,Cesaroni2005,Cesaroni2013,Cesaroni2014}.
In order to establish the disk accretion scenario as a viable route for the formation of the most massive stars \citep[e.g.,][]{Kuiper2010}, we seek
to simultaneously resolve the spatial morphology of the disk and the regions where the jet originates, and to determine whether or not the
disk gas is in centrifugal equilibrium. 

In the recent years, we have identified an ``isolated''  massive young star in the hot molecular core (HMC) of the star-forming region
G023.01$-$00.41, which is located about half way between the Sun and the Galactic center, at a parallax distance of 4.6\,kpc
\citep{Brunthaler2009}. The HMC emits a luminosity of 4\,$\times$\,$10^4$\,L$_{\odot}$ \citep{Sanna2014}, and stands out among
the strongest Galactic CH$_3$OH maser sources \citep{Menten1991,Cyganowski2009,Sanna2010,Moscadelli2011,Sanna2015}. The HMC
luminosity corresponds to that of a zero-age-main-sequence (ZAMS) star with a mass of 20\,M$_{\odot}$ and an O9 spectral-type 
\citep{Ekstrom2012}. The HMC is located at the center of a collimated CO outflow whose emission extends up to the parsec scales 
\citep{Araya2008,Furuya2008,Sanna2014}; we tracked the driving source of the outflow down to the HMC center, where we imaged a
collimated radio thermal jet associated with strong H$_2$O maser shocks \citep{Sanna2010,Sanna2016}. The different outflow tracers are
aligned on the sky plane and constrain the direction of the outflow axis within an uncertainty of a few degrees \citep{Sanna2016}. An
accurate knowledge of the outflow axis allows us to pinpoint the young stellar object (YSO) position, and thus to circumvent the usual
problem of distinguishing between velocity gradients due to expanding or rotational motions.


\begin{table*}
\caption{Summary of ALMA observations at Cycle\,3 (code 2015.1.00615.S).}\label{tobs}
\centering
\begin{tabular}{ c c c c c c c c c c}

\hline \hline
Array\,Conf. & R.A.\,(J2000) &        Dec.\,(J2000)    &   V$_{\rm LSR}$ & Freq.\,Cove. & $\Delta\nu$  &     BP\,Cal.        & Phase\,Cal. & Flux\,Cal. &  HPBW   \\
                   &      (h\,m\,s)   &  ($^{\circ}$\,$'$\,$''$) &   (km\,s$^{-1}$) &     (GHz)       &     (kHz)     &                        &                   &               &  ($''$)     \\
    (1)          &          (2)        &               (3)              &            (4)           &        (5)        &      (6)        &       (7)            &      (8)        &      (9)    &    (10)    \\
\hline
 & & & & & & & & & \\
C36--6 & 18:34:40.290 & --09:00:38.30 &  77.4  & 216.9, 236.5 & 488.3 & J1751$+$0939 & J1851$+$0035 & J1733$-$1304 & 0.20 \\
\hline
\end{tabular}
\tablefoot{Column\,1: granted 12m-array configuration. Columns\,2 and\,3: target phase center (ICRS system). Columns\,4: source radial velocity. Columns\,5:
minimum and maximum rest frequencies covered. The ALMA IF system was tuned at a LO frequency of 226.544\,GHz, and made use of 4 basebands evenly
placed in the lower and upper sidebands. Column\,6: maximum spectral resolution required. Columns\,7, 8, and\,9: bandpass, phase, and absolute flux calibrators
employed at both runs. Calibration sources were set by the ALMA operators at the time of the observations. Columns\,10: required beam size at a
representative frequency of 220.6\,GHz.}
\end{table*}


Here, we exploit the information about the star-outflow geometry in G023.01$-$00.41, and make use of Atacama Large Millimeter/submillimeter
Array, ALMA, observations (Sect.\,\ref{obs}), to directly resolve a disk-jet system in the vicinity of an O-type YSO. We firstly show that the line
emission from dense gas reveals a molecular disk which extends up to radii of 2000--3000\,au from the central star; the disk is warped in the
outer regions (Sect.\,\ref{res}). Then, we make use of the position-velocity (\emph{pv-}) diagrams of gas along the disk plane to show that
gas is falling in close to free-fall, and slowly rotating with sub-Keplerian velocities; at radii near 500\,au, gas rotation takes over, and could
approach centrifugal equilibrium at smaller radii, although higher-resolution observations are needed to study the inner regions (Sect.\,\ref{discus}).
In the process, we image the molecular jet component which arises from the inner disk regions. We corroborate our conclusions by comparing the
observed \emph{pv-}diagrams with those obtained for a disk model around a 20\,M$_{\odot}$ (Appendix\,\ref{model}).

\section{Observations and calibration}\label{obs}

We made use of our previous Submillimeter Array (SMA) observations at 1\,mm \citep{Sanna2014}, which covered a range
of angular resolutions between $3''$ and $0\farcs7$, to set up the requirements for higher resolution observations with ALMA.

We observed the star-forming region G023.01$-$00.41 with the 12\,m-array of ALMA in band 6 (211--275\,GHz).
Observations were conducted under program 2015.1.00615.S during two runs, on 2016 September 5 and 16 (Cycle\,3), 
with precipitable water vapor of 1.5\,mm and 0.6\,mm, respectively. The 12\,m-array observed with 45 antennas covering
a baseline range between 16\,m and 3143\,m, with the aim to achieve an angular resolution of $0\farcs2$, and to recover
extended emission over a maximum scale of $2''$. 

We made use of the dual-sideband receiver, in dual polarization mode, to record 12 narrow spectral windows, each 234\,MHz wide,
and an additional wide band of 1875\,MHz.  The narrow spectral windows were correlated with 960\,channels and Hanning smoothed by 
a factor of 2, achieving a velocity resolution\footnote{The CH$_3$CN\,($12_{K}$--$11_{K}$) lines were sampled with 
0.35\,km\,s$^{-1}$ spectral resolution.} of 0.7\,km\,s$^{-1}$. Individual spectral windows were placed to cover a number of molecular
lines of high-density tracers ($>$\,10$^6$\,cm$^{-3}$) such as methanol (CH$_3$OH) and methyl cyanide  (CH$_3$CN). These
settings provide a line spectral sampling of 13\,channels for an expected linewidth of 9\,km\,s$^{-1}$. The wide band was correlated
with 3840\,channels and centered at a rest frequency of 217.860\,GHz. This band was used to construct a pure continuum image from
the line-free channels. 

We spent 1\,hour of time on-source during a total observing time of 2\,hours, which includes calibration overheads. Time constraints 
were set to achieve a thermal rms per (narrow) spectral channel of about 2\,mJy\,beam$^{-1}$ (with 36 antennas), which corresponds to
a brightness temperature of 1.4\,K over a beam of $0\farcs2$. Additional observation information is summarized in Table\,\ref{tobs}. 

The visibility data were calibrated with the Common Astronomy Software Applications (CASA) package, version\,4.7.1 (r39339), making
use of the pipeline calibration scripts. We determined the continuum level in each spectral window separately. We made use of the pipeline
spectral image cubes and selected the line-free channels from a spectrum integrated over a circular box of $2''$ in size, which was centered
on the  target source. The task \emph{uvcontsub} of CASA was used to subtract a constant continuum level across the spectral window
(\emph{fitorder}\,$=$\,0).  We imaged the line and continuum emission with the task \emph{clean} of CASA. In each individual map, the
ratio between the image rms and the thermal noise, expected from the ALMA sensitivity calculator, is near unity. For the continuum map,
we integrated over a line-free bandwidth of 218\,MHz selected from the wide spectral window, and achieved a signal-to-noise ratio of about
100. Imaging information is summarized in Table\,\ref{tima}.


\begin{table*}
\caption{Imaging information.}\label{tima}
\centering
\begin{tabular}{c c r l c c c}
\hline \hline
 Tracer   &  $\nu$     &  E$_{\rm up}$  & Weight  &   HPBW         &        rms                 &  \textsc{S$_{\rm peak}$}  \\
             &   (GHz)     &        (K)            &             &  ($''$)       & mJy\,beam$^{-1}$ &   mJy\,beam$^{-1}$   \\
\hline
 & & & & & &  \\
\multicolumn{1}{l}{\textsc{Dust Continuum}}  & 217.8240  & ...  &   R0.5 &  \multicolumn{1}{c}{0.207}  & 0.16   & 14.63 \\
\multicolumn{1}{l}{CH$_3$OH\,($4_{2,2}$--$3_{1,2}$)\,E}  & 218.4400  &  45.4  &   R0.5    &  \multicolumn{1}{c}{0.184}  & 3.4\,$^{a}$  & 220.1$^{a}$  \\
\multicolumn{1}{l}{CH$_3$OH\,($10_{2,8}$--$9_{3,7}$)\,A$^{+}$}        & 232.4186  &  165.4  &   R0    &  \multicolumn{1}{c}{0.155}  & 1.0  & 114.4  \\
\multicolumn{1}{l}{CH$_3$OH\,($18_{3,16}$--$17_{4,13}$)\,A$^{+}$}  & 232.7835  &  446.5  &   R0    &  \multicolumn{1}{c}{0.155}  & 1.0  & 107.3  \\
\multicolumn{1}{l}{CH$_3$CN\,($12_{3}$--$11_{3}$)}  & 220.7090  &  133.1  &   R0    &  \multicolumn{1}{c}{0.155}  & 1.4  & 110.4 \\
\multicolumn{1}{l}{CH$_3$CN\,($12_{4}$--$11_{4}$)}  & 220.6793  &  183.1  &   R0    &  \multicolumn{1}{c}{0.155}  & 1.4  & 98.1 \\
\hline
\end{tabular}
\tablefoot{Columns\,1 and\,2 list the tracer and central frequency of each map (or molecular transition), respectively. For the line
emission, column\,3 specifies the upper excitation-energy of the molecular transition. Column\,4 reports the Briggs' robustness
parameter set for the imaging. Column\,5 reports the restoring (circular) beam size, set equal to the geometrical average of the
major and minor axes of the dirty beam size. Columns\,6 and\,7 report the rms of the map and the peak brightness of the
emission, respectively. $^{a}$\,Units of mJy\,beam$^{-1}$\,km\,s$^{-1}$.}
\end{table*}


\begin{figure*}
\centering
\includegraphics [angle= 0, scale= 1.0]{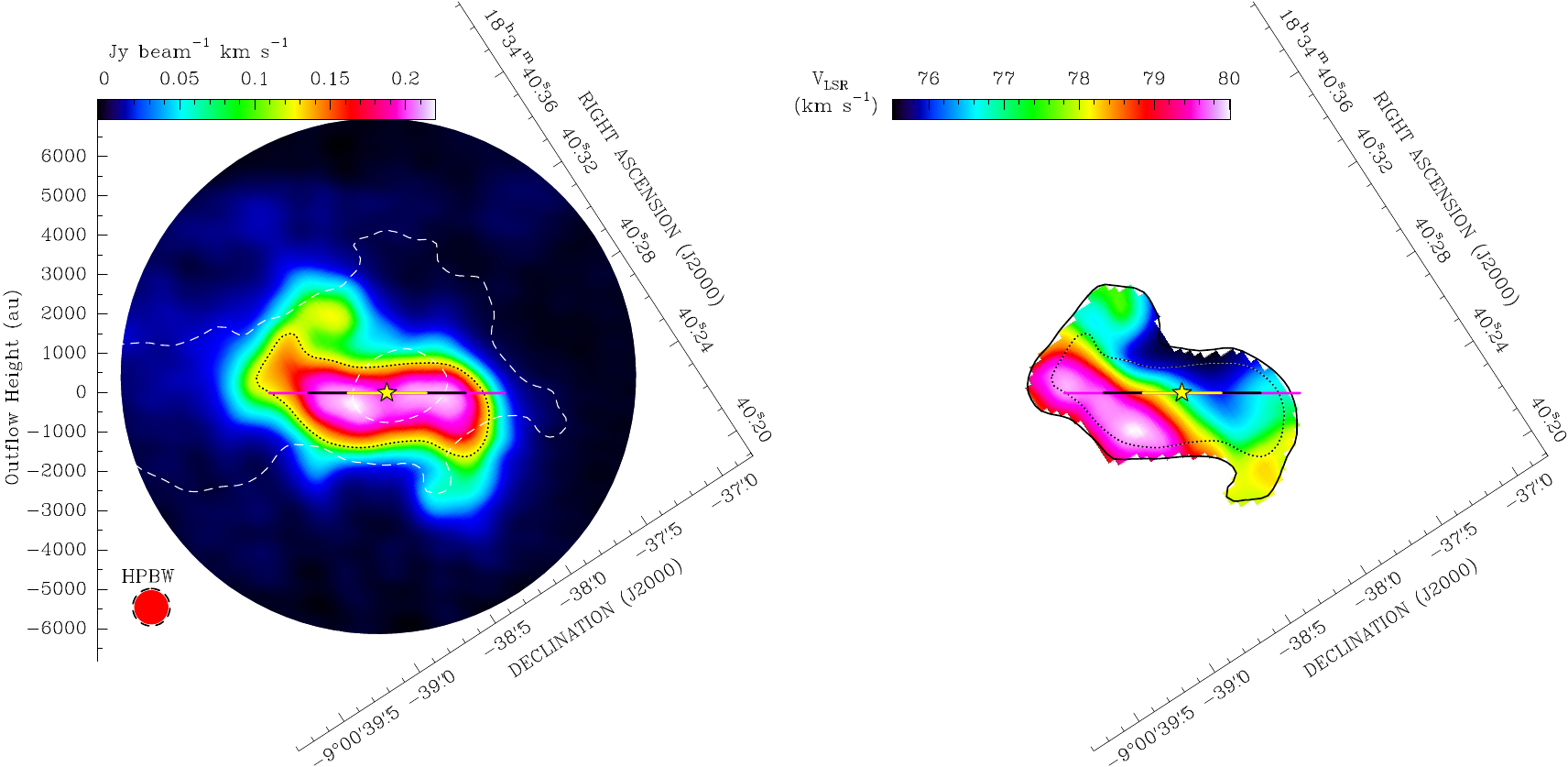}
\caption{Kinematic analysis of a low excitation-energy line (E$_{\rm up}$ of 45\,K) of the methanol gas emission towards G023.01$-$00.41. 
\textbf{Left:} moment-zero map of the CH$_3$OH\,($4_{2,2}$--$3_{1,2}$)\,E line emission (colors) combined with the continuum map of the dust emission
at 1.37\,mm (dashed white contours). Maps have been rotated clockwise by a position angle of $-57^{\circ}$, in order to align the (projected) outflow axis, drawn
on the left side, with the north-south axis; negative outflow heights indicate the receding outflow direction. The CH$_3$OH emission was integrated in the range
78.8--80.2\,km\,s$^{-1}$; the wedge on the top left corner quantifies the line intensity, from its peak to the maximum negative in the map. The lowest dashed
contour corresponds to the 10\,$\sigma$ level of the dust map, and the inner contour traces the 50\% level from the continuum peak emission. The disk plane
is drawn at three radii: from the central star position  (star) to 1000\,au (yellow), from 1000 to 2000\,au (black), and up to 3000\,au (pink). The dotted black contour
draws the 60\% level of the CH$_3$OH emission, which identifies the disk profile (see text). The synthesized ALMA beams, for the dust continuum map (dashed
circle) and the line map (red circle), are shown in the bottom left corner.
\textbf{Right:} first-moment map (colors) of the CH$_3$OH\,($4_{2,2}$--$3_{1,2}$)\,E line emission plotted in the left panel. The outer contour 
traces the 40\% level of the moment-zero map; the inner dotted contour is the same as in the left panel. The LSR velocity scale is drawn in the upper left.} 
\label{mom}
\end{figure*}

\begin{figure}
\centering
\includegraphics [angle= 0, scale= 0.5]{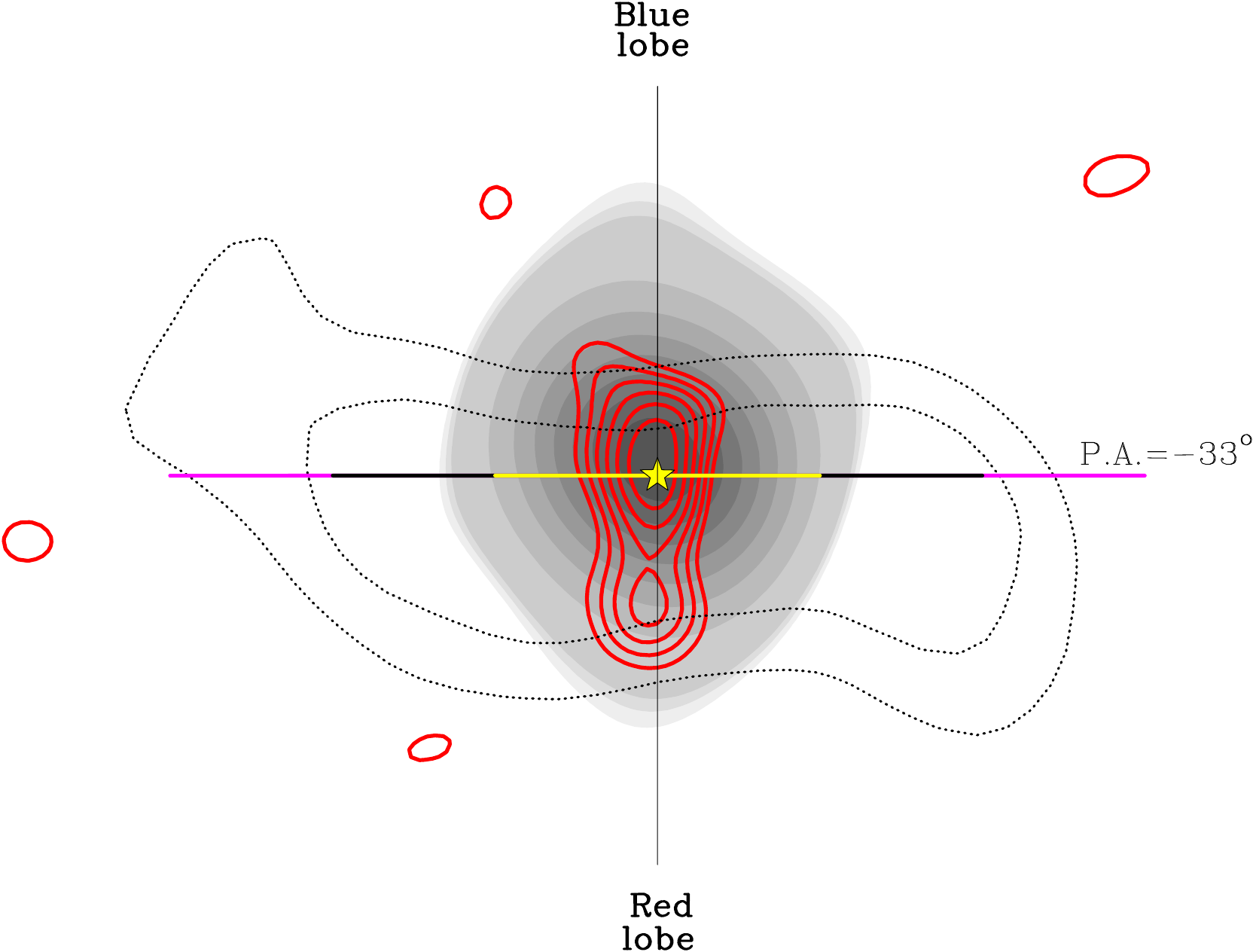}
\caption{Comparison between the disk emission imaged in Fig.\,\ref{mom}  (dotted profile) and the radio continuum emission at 22\,GHz (grey) and 45\,GHz (red) 
detected towards  G023.01$-$00.41. Dotted contours mark the 60\% and 80\% levels of the CH$_3$OH\,($4_{2,2}$--$3_{1,2}$)\,E line
emission showing the disk morphology. The disk plane at a position angle of $-33^{\circ}$ and the star position are defined as in  Fig.\,\ref{mom}. The radio continuum 
emission imaged with the Karl G. Jansky Very Large Array, at a similar resolution as the ALMA observations, traces a radio thermal jet \citep[from][]{Sanna2016}.
The lower three contours of the 22\,GHz emission start at 5\,$\sigma$ by 1\,$\sigma$ steps of 8\,$\mu$Jy\,beam$^{-1}$, and then increase at steps of
10\,$\sigma$. The 45\,GHz contours start at 3\,$\sigma$ by 1\,$\sigma$ steps of 22\,$\mu$Jy\,beam$^{-1}$. The outflow direction perpendicular to the 
disk plane is indicated by a black line, and the orientation of the blueshifted and redshifted outflow lobes is also specified. The linear scale of the image is quantified by
the colored ticks along the disk plane (see Fig\,\ref{mom}).} 
\label{diskradio}
\end{figure}


\begin{figure*}
\centering
\includegraphics [angle= 0, scale= 0.47]{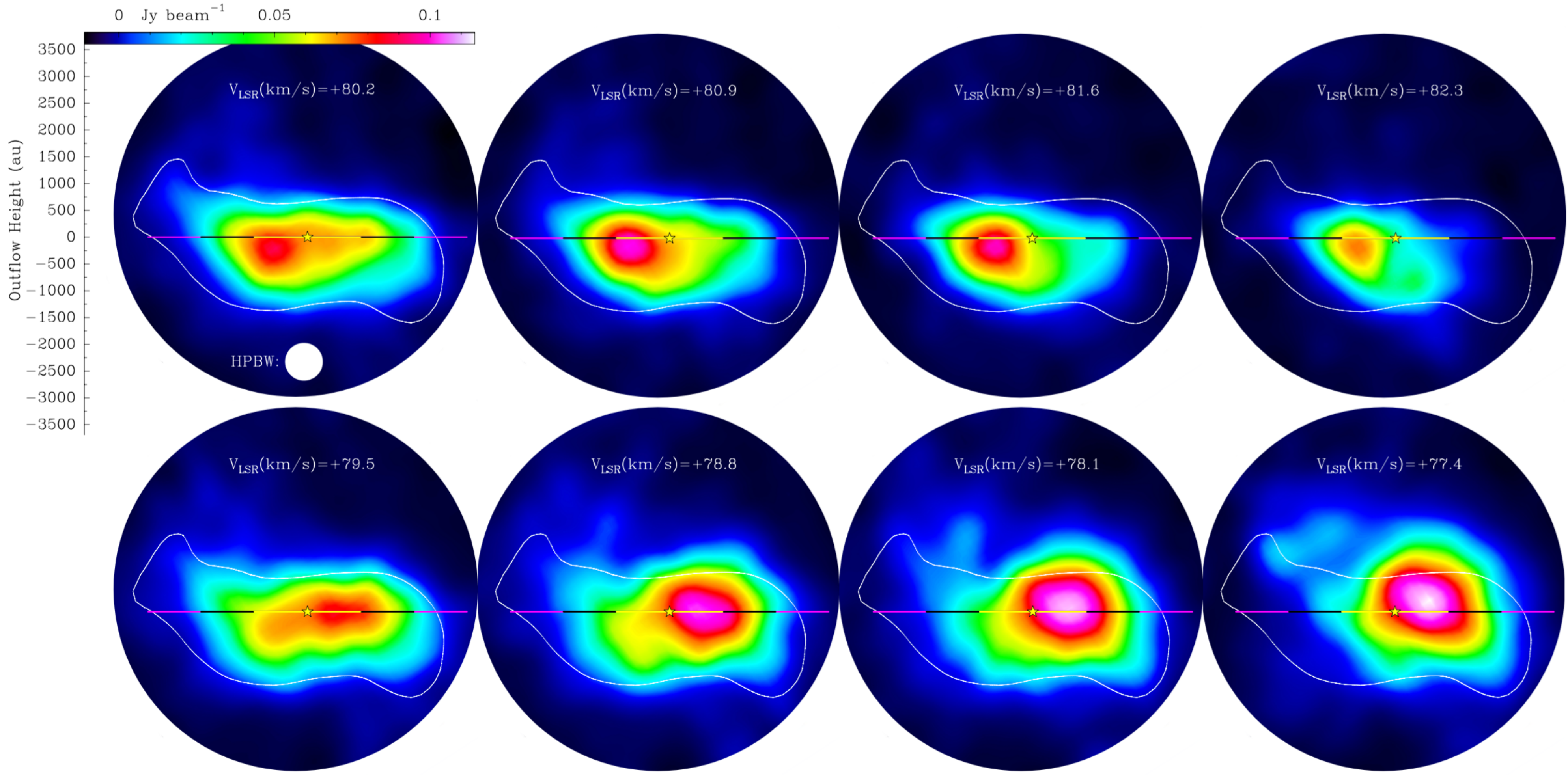}
\caption{Channel maps of the CH$_3$OH\,($10_{2,8}$--$9_{3,7}$)\,A$^{+}$ line emission (colors) observed at 700\,au resolution with
ALMA (synthesized beam in white). Each map is labeled by its central velocity (V$_{\rm LSR}$). The brightness scale of the line emission is
quantified by the wedge on the top left panel. The reference system and symbols in each channel map are the same used in Fig.\,\ref{mom}.
For comparison, the white contour corresponds to the dotted black contour of Fig.\,\ref{mom}.} 
\label{chmap165}
\end{figure*}


\section{Results}\label{res}

According to the ``observer's definition'' of disk framed in \citet{Cesaroni2007}, necessary conditions for a disk detection are \textbf{(i)} a
flattened core of gas and dust perpendicular to the outflow direction, and \textbf{(ii)} a velocity gradient along its major axis. We therefore
want to analyze the spatial morphology and kinematics of dense gas which lies at the center of the bipolar outflow, and, specifically, along a
position angle of $-33^{\circ}$ on the sky plane (measured east of north). This direction traces the projection of the equatorial plane, hereafter
the ``disk plane'', perpendicular to the molecular outflow and the radio jet orientation (or, simply, the outflow), which is inclined by less than
$30^{\circ}$ with respect to the sky plane \citep{Sanna2014,Sanna2016}.

In order to gain an overall view of the mass distribution around the star, we first looked for a molecular tracer of cold and dense gas
($>$\,10$^6$\,cm$^{-3}$). In the left panel of Fig.\,\ref{mom}, we plot the velocity-integrated map of the CH$_3$OH gas emission at a
frequency of 218.440\,GHz (in color); this CH$_3$OH transition has an upper excitation-energy of only 45\,K (E$_{\rm up}$). To emphasize
the bulk of the emission, we integrated over a small velocity range (2\,km\,s$^{-1}$) near the line peak.  

\begin{figure*}
\centering
\includegraphics [angle= 0, scale= 0.6]{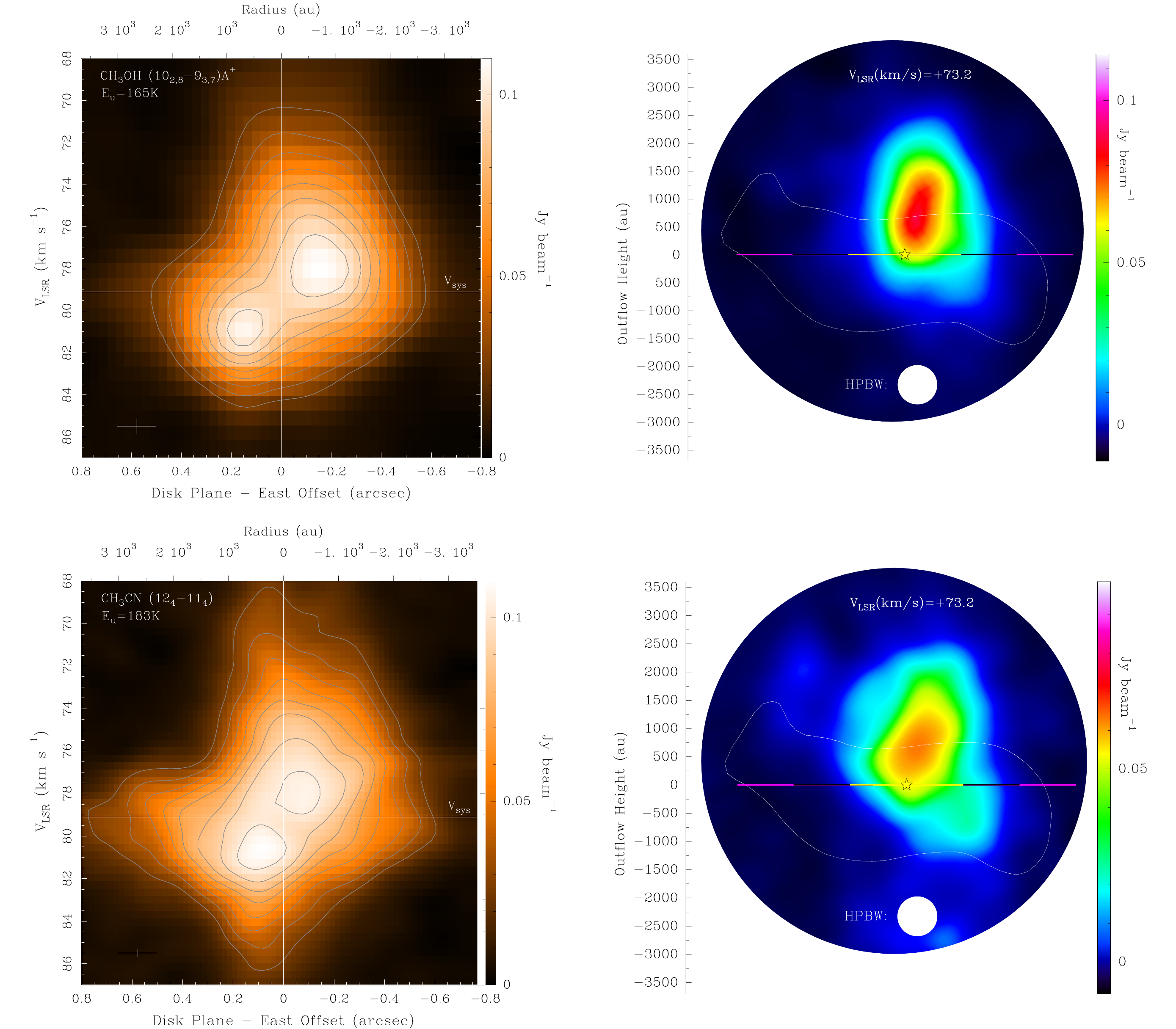}
\caption{Velocity field analysis of the CH$_3$OH  (upper) and CH$_3$CN (lower) line emission along the disk plane. 
\textbf{Left column:} \emph{pv}-diagrams of gas lying along the disk plane (cut at $-33^{\circ}$). Molecular labels are on the upper left corner
of each panel. East offsets are measured along the disk plane starting from the YSO position; the same scale in astronomical units is written on
the upper axis. For each panel, contours start at 90\% of the peak emission and decrease by 10\% steps; colors are drawn according to the
wedges on the right sides. The rest velocity of the star is set at $+79.1$\,km\,s$^{-1}$. The spatial and spectral resolutions are indicated in the
bottom left corners. The CH$_3$CN observations have two times higher spectral resolution than those of the CH$_3$OH line.
\textbf{Right column:} channel maps of the blueshifted jet component, at a velocity of $+73.2$\,km\,s$^{-1}$, which contributes to 
the high blueshifted velocities in the \emph{pv}-diagrams on the left. Symbols as in Figs.\,\ref{mom} and\,\ref{chmap165}.} 
\label{pv}
\end{figure*}

For clarity, we have rotated this map in order to align the outflow axis, which lies at a position angle of $+57^{\circ}$ on the sky plane
\citep{Sanna2016}, with the vertical axis of the plot. The blueshifted (approaching) lobe of the outflow points now to the north. The horizontal
axis is aligned with the disk plane, which is marked in the plot at multiples of 1000\,au from the central star position. The position of the central
star was set as to maximize the symmetry of the blue- and redshifted sides of the disk (see Figs.\,\ref{chmap165} and~\ref{pv}), and has
coordinates of R.A.\,(J2000)\,$=$\,18$^{\rm h}$34$^{\rm m}$40$.^{s}$283 and Dec.\,(J2000)\,$=$\,--9$^{\circ}$00$'$38$''.$310 
($\pm$\,30\,mas per coordinate). This position also coincides with the peak position of the radio continuum emission observed with the Very
Large Array at the same resolution as the ALMA beam \citep{Sanna2016}. The same geometry and symbols will be used in all maps for comparison.

Figure\,\ref{mom} shows that dense gas condenses in the direction perpendicular to the outflow axis. The CH$_3$OH gas emission extends
up to a deconvolved radius of 2380\,$\pm$\,840\,au (or 520\,mas) from the central star, and has a ratio between the minor-to-major
axes of 0.2. These values are calculated by Gaussian fitting the CH$_3$OH emission within the 60\% contour level (dotted black contour in
Fig.\,\ref{mom}). We use this threshold to select the region where the CH$_3$OH iso-contours have a nearly constant height. \emph{This
piece of evidence satisfies the first condition, that of a highly flattened structure perpendicular to the outflow,} and hereafter we refer to the
60\% contour level as the ``disk profile''. Below the 60\% contour level, the CH$_3$OH emission emerges  at (relatively) large radii from the
star ($>$2000\,au) and appears significantly warped with respect to the disk plane. In this paper, we focus on the analysis of the emission
within the disk profile.

In the right panel of Fig.\,\ref{mom}, we plot the intensity-averaged map of the velocity field of the CH$_3$OH gas  (i.e., the first-moment 
map). There is a clear gradient between the redshifted and blueshifted velocities moving from the left to the right side of the star along the 
disk plane. \emph{This second piece of evidence fulfills the more stringent requirement for confirming a disk candidate.} On a closer look,
the velocity field is actually skewed along the main outflow direction. This is not surprising, since, around Solar-mass stars, CH$_3$OH
emission has been shown to be excited both in the disk and jet regions \citep[e.g.,][]{Leurini2016,Lee2017c,Bianchi2017}. Indeed, contamination
of the velocity gradient  across the disk plane, by the blueshifted and redshifted gas velocities due to the outflow emission, to the north
and south of the star respectively, can explain this apparent velocity gradient.

We stress that the physical structure imaged in Fig.\,\ref{mom} (left) does not coincide with the definition of ``toroid'', which is a
non-equilibrium structure with a mass comparable or greater than that of the central star(s) and a size of the order of 10,000\,au \citep{Cesaroni2005b,Beltran2005}. 
The rotating toroid surrounding the massive YSO in G023.01$-$00.41 was previously imaged and discussed in \citet{Furuya2008} and \citet{Sanna2014}.
In Fig.\,\ref{diskradio}, we directly compare the relative orientation of the disk profile with that of the radio thermal jet emission detected at  
22 and 45\,GHz by \citet{Sanna2016}, showing that they are clearly perpendicular.

To better investigate the gas kinematics inside the disk profile, in Fig.\,\ref{chmap165} we plot a number of maps of the
CH$_3$OH\,($10_{2,8}$--$9_{3,7}$)\,A$^{+}$  line emission at different velocities. This transition has an excitation-energy
(E$_{\rm up}$\,$=$\,165\,K) about four times higher than that of the line in Fig.\,\ref{mom}. The upper row shows the spatial morphology
of the CH$_3$OH gas emission at increasing (redshifted) LSR velocities, from left to right, and underlines the eastern disk side. On
the contrary, the lower row shows the CH$_3$OH gas distribution at decreasing (blueshifted) LSR velocities, which arise from the western
side of the disk. Each row spans a velocity range of 2.8\,km\,s$^{-1}$, which is the rotation speed expected around a 20\,M$_{\odot}$
star at an outer radius of 2300\,au (assuming centrifugal equilibrium). 

Figure\,\ref{chmap165} resolves the spatial distribution of gas around the star at the different velocities, providing direct proof that the CH$_3$OH
gas is rotating clockwise around the outflow axis (as seen from the north), with the approaching and receding sides of the disk to the west
and east of the central star, respectively. The pair of maps on the same column to the left shows the transition between the eastern (upper)
and western (lower) sides of the disk emission, and can be used to infer the rest velocity of the star (see below). Between velocities of 80
and 79\,km\,s$^{-1}$, the CH$_3$OH gas emission is maximally stretched along the disk plane on either sides of the star; at higher (lower)
velocities, the emission progressively brightens southeastward (northwestward) of the disk plane, in agreement with the redshifted (blueshifted)
outflow lobe. A posteriori, this behavior confirms that the observed molecular species can trace both the disk kinematics and the (inner)
outflowing gas (see also Fig.\,\ref{pv}, right).

\onlfig{
\begin{figure*}
\centering
\includegraphics [angle= 90, scale= 0.50]{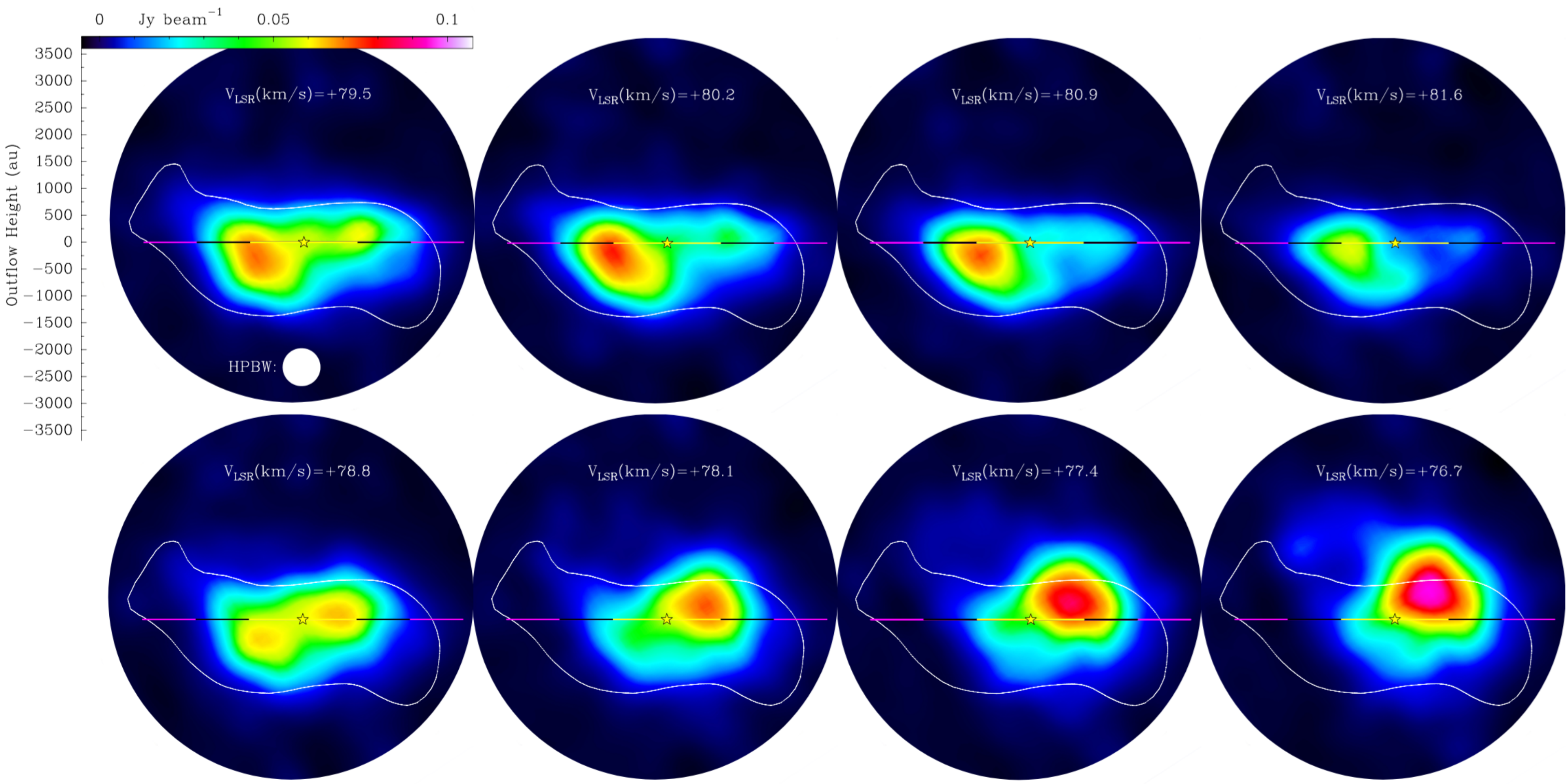}
\caption{Similar to Fig.\,\ref{chmap165}, but for the CH$_3$OH\,($18_{3,16}$--$17_{4,13}$)\,A$^{+}$ line emission with E$_{\rm up}$ of
446\,K. Note that there is a shift of $-0.7$\,km\,s$^{-1}$ with respect to the channel maps in Fig.\,\ref{chmap165}.} 
\label{chmap446}
\end{figure*}
}

In the Fig.\,\ref{chmap446}, we provide the same analysis for the CH$_3$OH\,($18_{3,16}$--$17_{4,13}$)\,A$^{+}$ line emission, which
has an excitation-energy of 446\,K  (E$_{\rm up}$), and is more affected by the outflow emission. For this line, the transition between
the red- and blueshifted sides of the disk occurs at lower velocity than in Fig.\,\ref{chmap165}. The rest velocity of the star, V$_{\star}$, was estimated
from the median value between Fig.\,\ref{chmap165} and Fig.\,\ref{chmap446}, and is set to $+79.1$\,km\,s$^{-1}$ with an uncertainty of
$\pm$\,0.4\,km\,s$^{-1}$. We explicitly note that, since the disk plane is not seen exactly edge-on, the blueshifted outflow emission close to
the star, which lies between the observer and the disk, shifts the velocity field towards the lower velocities, when weighting the velocity channels
by the line intensity. In the first-moment map of Fig.\,\ref{mom}, the net result is that the redshifted velocities are down-weighted, and the central
velocity is shifted by about $-1.5$\,km\,s$^{-1}$ with respect to the rest velocity of the star. 

Overall, in Figs.\,\ref{mom} and~\ref{chmap165} we resolve the gas emission within a few 1000\,au of the O-type YSO, which allows us to prove
the presence of a disk, and prompts us to study its kinematics in detail.

\section{Discussion}\label{discus}

In the following, we want to study the dependence of the gas velocity on the distance to the star, and to quantify the disk mass. Together with
the observations, we discuss the results of a radiation transfer model for a circumstellar disk around a 20\,M$_{\odot}$ star (Appendix\,\ref{model}).
We have simulated the disk appearance for our specific observational conditions, and produced synthetic maps of the line and dust emission around the
young star, in order to compare the observed \emph{pv}-diagrams with those expected under two simple assumptions: the disk is falling in towards
the central star due to gravitational attraction; the disk is rotating around the central star in centrifugal equilibrium.

\subsection{Position-velocity diagrams}

In Fig.\,\ref{pv}, we study the velocity profile of gas along the disk plane through the \emph{pv-}diagrams of two molecular species having a common
methyl group, CH$_3$OH and CH$_3$CN. On the same column to the left, we compare the \emph{pv-}diagrams of the 
CH$_3$OH\,($10_{2,8}$--$9_{3,7}$)\,A$^{+}$ and CH$_3$CN\,($12_{4}$--$11_{4}$) line transitions, which have similar excitation-energies of
165\,K and 183\,K, respectively. The two plots span the same ranges in space and velocity, and the lower intensity contour is drawn as to include an
outer radius of 2400\,au. Both \emph{pv-}diagrams are double peaked, and the peaks are symmetrically displaced with respect to the star position and
velocity. However, the CH$_3$CN line emission shows a steeper velocity gradient at small offsets from the star, which indicates that this transition 
traces a region closer to the disk center than the CH$_3$OH transition. This piece of evidence also explains the lower redshifted tail of the CH$_3$CN
\emph{pv-}diagram, which traces the higher rotation speeds approaching the central star, as it can be expected for Keplerian-like rotation
($v_{\rm rot}\propto R^{-1/2}$). This component is missing in the roundish contours of the CH$_3$OH emission. In Fig.\,\ref{pvK3}, we plot the
\emph{pv-}diagram of the CH$_3$CN\,($12_{3}$--$11_{3}$) line, which  has an upper excitation-energy of 133\,K, and shows a similar profile to that
of the K\,$=$\,4 line. Their agreement, when compared to the \emph{pv-}diagram of the CH$_3$OH line, strengthens the idea that the CH$_3$CN gas
allows us to peer into the innermost disk regions, and this holds for a broad range of line excitation-energies. 
 
\onlfig{
\begin{figure*}
\centering
\includegraphics [angle= 0, scale= 0.45]{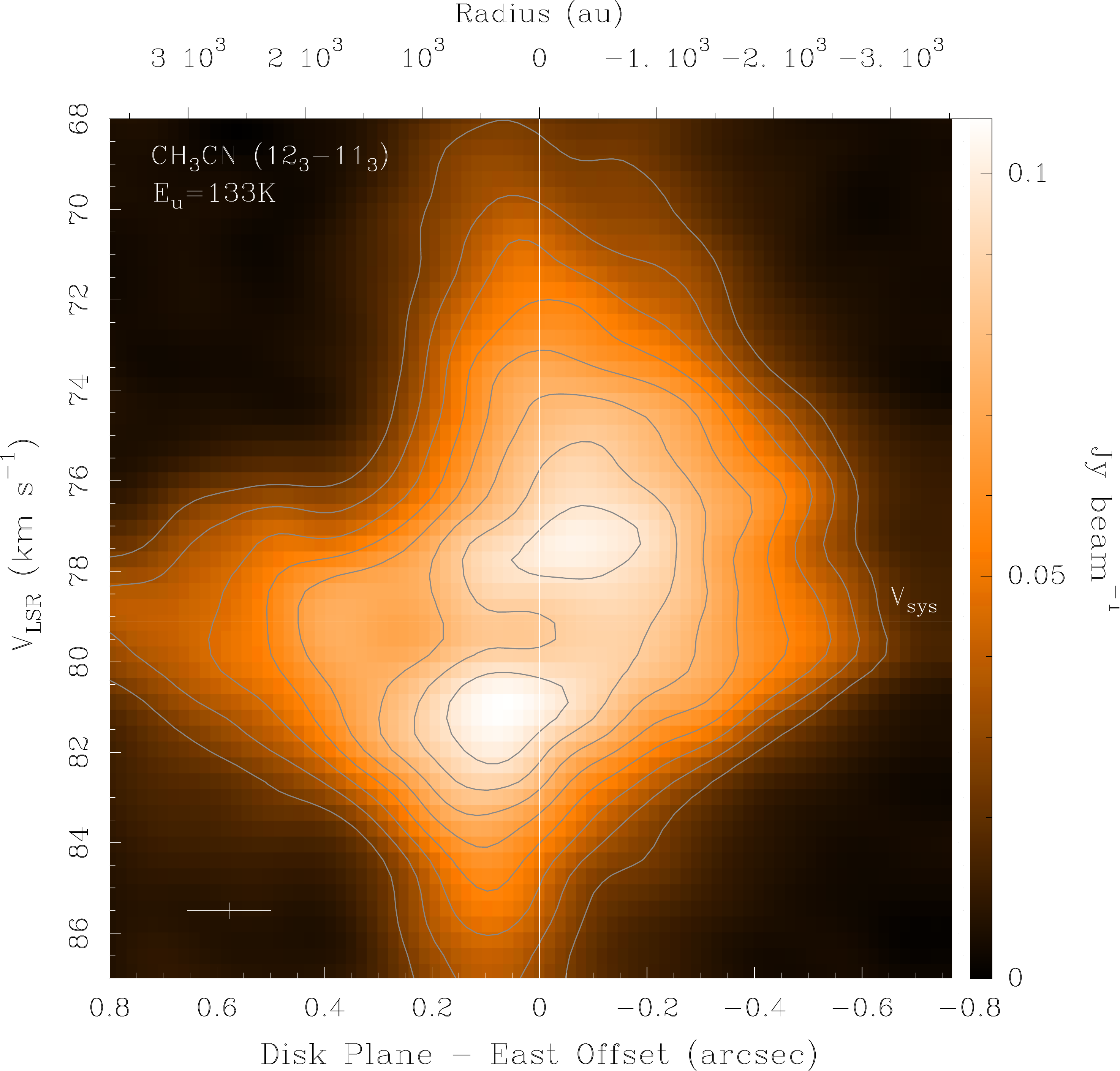}
\caption{\emph{pv}-diagram of the CH$_3$CN\,($12_{3}$--$11_{3}$) line emission, to be compared with that
of the CH$_3$CN\,($12_{4}$--$11_{4}$) line in Fig.\,\ref{pv}. Same limits and symbols.} 
\label{pvK3}
\end{figure*}
}

\begin{figure*}
\centering
\includegraphics [angle= 0, scale= 1.0]{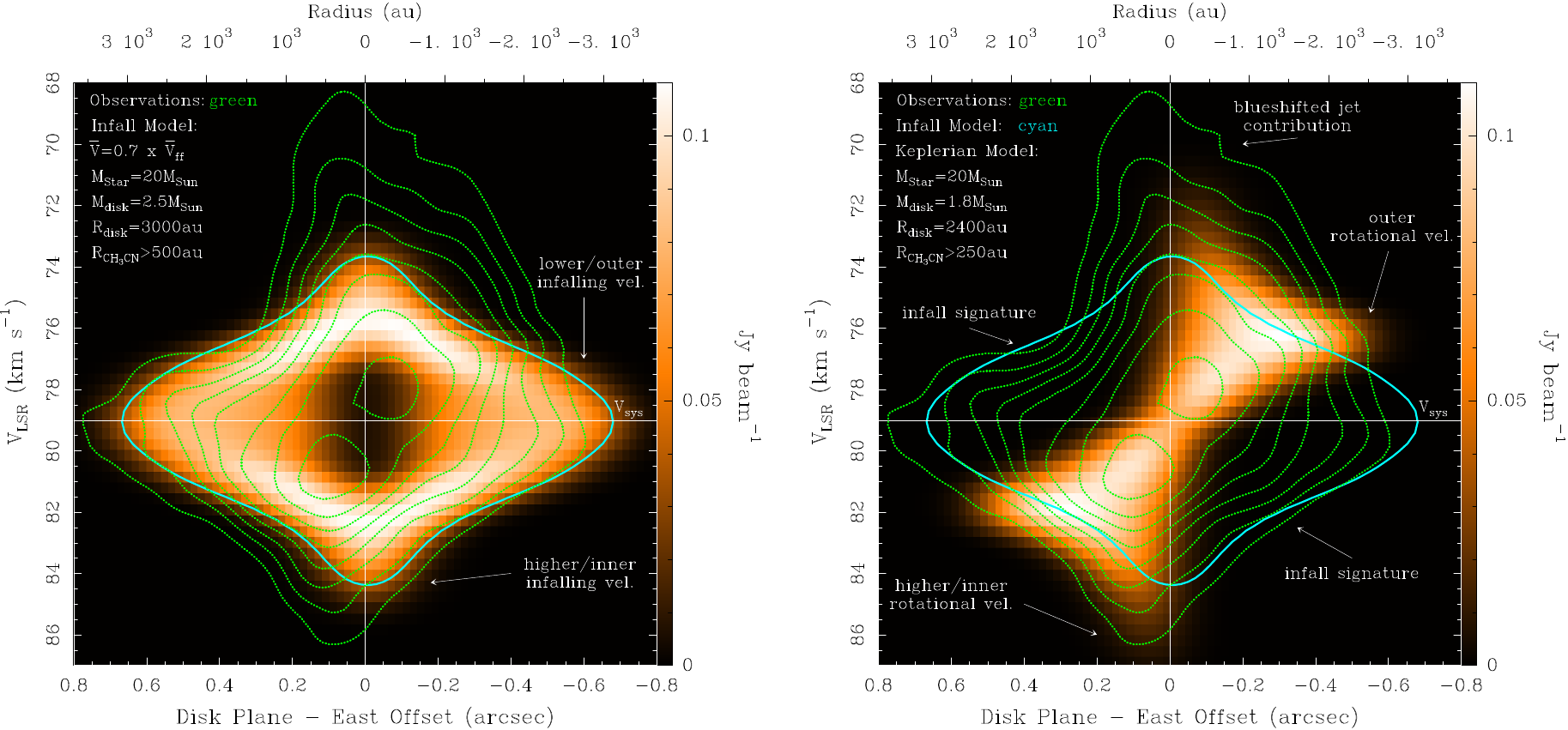}
\caption{Modeled \emph{pv}-diagrams of the CH$_3$CN\,($12_{4}$--$11_{4}$) line emission (color scale) along the plane of a circumstellar
disk around a 20\,M$_{\odot}$ star. The disk is inclined by $10^{\circ}$ with respect to the line-of-sight. 
\textbf{Left panel:} \emph{pv}-diagram obtained under the assumption of an infalling velocity field between radii of 3000\,au and 500\,au from
the central star. At each point, the magnitude of the velocity vectors is $70\%$ of the free-fall velocity. The plot limits and symbols are the same
used in Fig.\,\ref{pv}. For comparison, the green contours are those plotted for the same CH$_3$CN line in Fig.\,\ref{pv}; the cyan (modeled) contour
marks the same absolute level of the outer observed isocontour. This infalling profile well reproduces the observed velocities in the second and fourth
quadrants, and represents clear evidence of an accretion flow through the disk. 
\textbf{Right panel:} similar to the left panel, but for a purely Keplerian motion. The gas emission now extends between radii of 2400\,au and
250\,au from the central star. The cyan contour is the same as in the left panel. This rotational profile well reproduces the higher redshifted velocities
detected in the third quadrant (lower green contours). The green contours also match the outer rotational velocities in the first quadrant.} 
\label{pvmod}
\end{figure*}

On the other hand, both the CH$_3$CN and CH$_3$OH lines show increasing blueshifted velocities moving close to the star. In the right column of
Fig.\,\ref{pv}, we show a channel map at the blueshifted velocity of +73.2\,km\,s$^{-1}$ for both transitions. These maps clearly show a compact
(molecular) outflow component, namely the jet, to the north of the star, corresponding to the direction of the blueshifted outflow lobe. Since the
blueshifted outflow emission lies in between the observer and the disk, we cannot neglect its influence on the \emph{pv}-diagrams (see below).
Theory predicts that jets are launched and collimated in the inner few 100\,au from the central star \citep[e.g.,][]{Frank2014,Kuiper2015,Kuiper2016}. 
The jet width of 1510\,$\pm$\,73\,au (or 329\,mas), determined from the deconvoled (Gaussian) size of the CH$_3$OH emission, allows us to
set an upper limit of 800\,au to the radius where the jet originates. Notably, we do not detect the redshifted lobe of the jet, which lies in the
background with respect to the disk, and interpret this result as evidence for the inner disk regions being (partially) optically thick at
1\,mm \citep[e.g.,][]{Sanna2014,Forgan2016}.

In Fig.\,\ref{pvmod}, we compare the observed \emph{pv}-diagram of the CH$_3$CN\,($12_{4}$--$11_{4}$) line, which is sensitive to the inner disk
velocities, with the modeled \emph{pv}-diagrams for two different velocities fields. On the left panel, we consider an infalling disk around a 20\,M$_{\odot}$
star (colors), where the gas is moving radially at $70\%$ of the corresponding free-fall velocity ($v_{\rm ff} = 2GM_{\star}/R^{1/2}$). The disk extends
from the dust sublimation radius, where the disk temperature approaches 1500\,K, up to a radius of 3000\,au from the central star. We find that, this
fraction of the free-fall velocity, combined with an inner cutoff near 500\,au, best match the velocity range covered by the observed \emph{pv}-diagram
(green contours). Note that, since the infalling motion starts with non-zero velocities at the outer radius, this produces the inner hole near zero offsets in
the modeled \emph{pv}-diagram. The infalling profile well reproduces the observed \emph{pv}-diagram in the second and fourth quadrants, which are the
quadrants forbidden under the assumption of purely rotational motion (right panel). 

In the right panel of Fig.\,\ref{pvmod}, we show the comparison of the observed \emph{pv}-diagram (green contours) with that expected for a disk in
Keplerian rotation around a 20\,M$_{\odot}$ star (colors). We also draw the outer contour of the infalling disk model (cyan contour), in order to highlight
the regions excluded by simple rotation. At variance with the infalling profile, we find that a Keplerian-like profile better reproduces the higher velocities 
in the first and third quadrants, when we assume an inner cutoff near 250\,au and outer disk radius of 2400\,au, similar to the deconvolved size of
our disk profile (Fig.\,\ref{mom}).

This analysis shows that the velocity field of gas through the disk is a combination of infalling and rotational motions, within a radius of 3000\,au
from the central star. On the one hand, the detection of a radial flow towards the central star implies that the condition of centrifugal equilibrium does
not hold outside a radius of 500\,au, where the velocity field has to be a combination of sub-Keplerian rotation and infalling motion. This scenario 
resembles that of the sub-Keplerian infalling disks modeled by \citet{Seifried2011} under the presence of strong magnetic fields. Our
model predicts a mass infall rate at a radius of 500\,au of $6\times10^{-4}$\,M$_{\odot}$\,yr$^{-1}$. On the other hand, our data suggest that
the inward gas flow slows down in the inner disk regions ($\leq$\,500\,au), where the \emph{pv}-diagram could resemble that of a centrifugally
supported disk (see the right panel of Fig.\,\ref{pvmod}). However, at the moment this possibility is quite speculative, as the inner regions are sampled by
only two beams with the current resolution. We explicitly note that, although we fixed the star mass in our model, the simple agreement between observed
and modeled \emph{pv}-diagrams provides indirect confirmation of the central mass determined from the bolometric luminosity.


\begin{figure*}
\centering
\includegraphics [angle= 0, scale= 0.42]{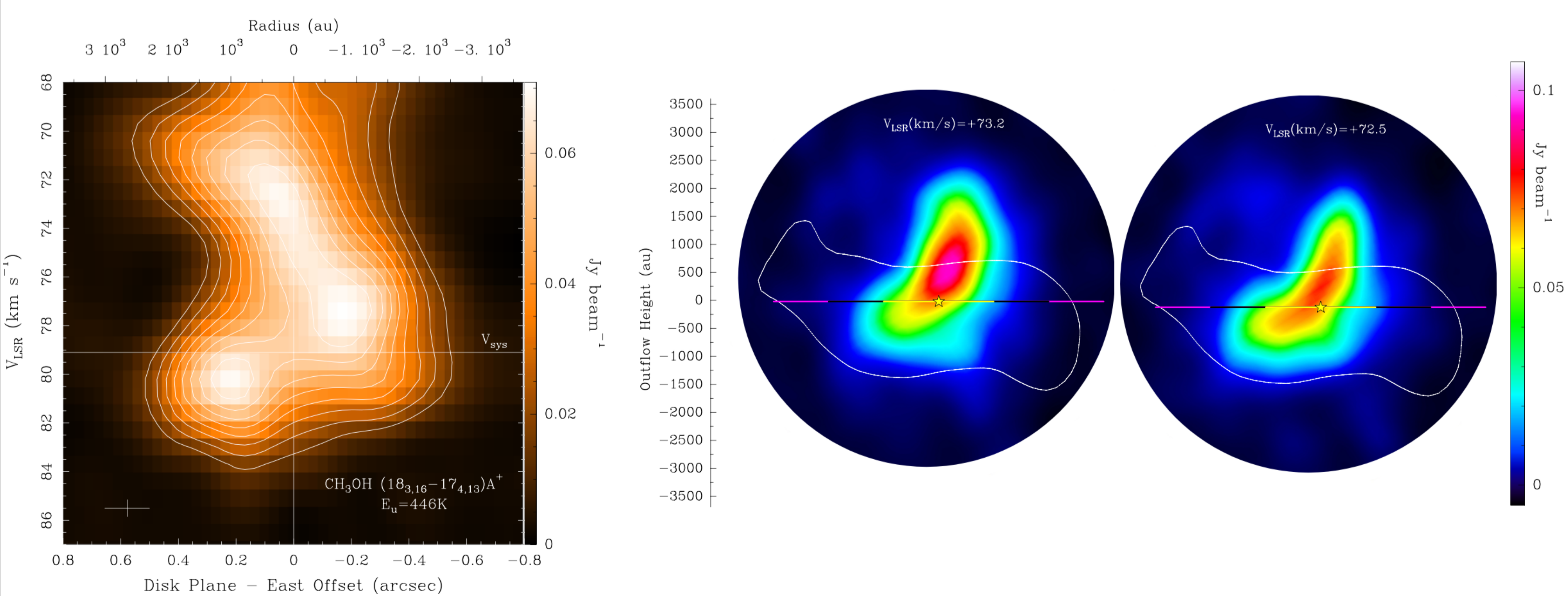}
\caption{Similar to Fig.\,\ref{pv}, but for the CH$_3$OH\,($18_{3,16}$--$17_{4,13}$)\,A$^{+}$ line emission with E$_{\rm up}$
of 446\,K. The right panels show two channel maps of the jet emission at a V$_{\rm LSR}$ of $+73.2$ (middle) and $+72.5$\,km\,s$^{-1}$
(right).} 
\label{pv446}
\end{figure*}


Moreover, in the right panel of Fig.\,\ref{pvmod} it is evident that there is an excess of blueshifted emission near zero offsets. This emission coincides
with the molecular jet detected in our channel maps (right column of Fig.\,\ref{pv}). If the molecular species does not trace the disk gas exclusively, and
the disk plane is not seen edge-on, we show that the blueshifted outflow emission close to the star leaves a strong imprint in the \emph{pv}-diagrams. 
In the Fig.\,\ref{pv446}, we provide the same \emph{pv}-analysis for the CH$_3$OH\,($18_{3,16}$--$17_{4,13}$)\,A$^{+}$ line emission. Interestingly,
at the higher excitation-energy of this line, the blueshifted jet emission strongly affects the eastern side of the \emph{pv}-diagram. In the channel maps
(right panels), the spatial distribution of the jet emission bends to the redshifted side of the disk indeed, and progressively aligns to the outflow axis
moving away from the disk plane. These maps definitely underline that the jet emission arises in the inner disk regions ($<800$\,au), providing a 
mechanism to transfer angular momentum away from the disk, and ensuring an inward flow of mass towards the central star.  

Under steady state accretion, the mass infall rate of our model would imply that the final star mass could be as large as three times the 
current value, assuming the accretion phase lasts for about 10$^{5}$\,yr. Nevertheless, it has been recently shown that the accretion process
of young massive stars undergoes episodic accretion bursts, with the accretion rate suddenly rising by a few orders of magnitude 
\citep{Caratti2017,Hunter2017}. Therefore, the net mass accretion onto the central star might exceed that inferred from the model.


\begin{figure}
\centering
\includegraphics [angle= 0, scale= 0.3]{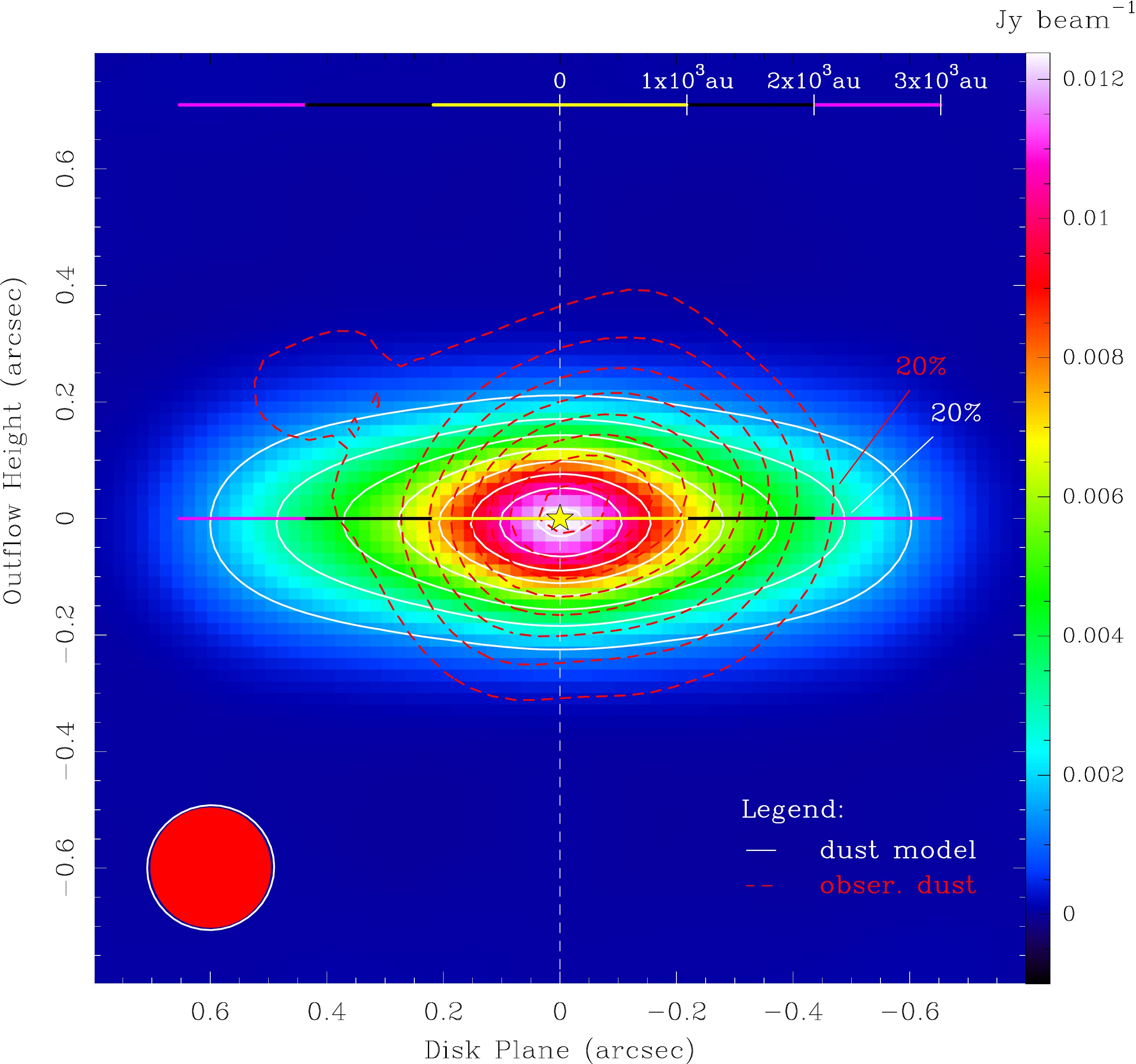}
\caption{Analysis of the dust continuum emission from the disk. 
Comparison of the dust emission observed with ALMA towards G023.01$-$00.41 (red dashed contours) with the 
modeled dust emission from a circumstellar disk around a 20\,M$_{\odot}$ star (colors and white contours). The brightness
scale of the modeled emission is quantified by the righthand wedge. The red dashed contours are plotted at steps of 10\% of
the observed peak emission, starting from 90\%; the white contours draw the same absolute levels for the model. The reference
system and symbols are the same used in the previous figures. The synthesized ALMA beams, for the observed (red filled circle) and
modeled (white circle) dust continuum maps, are shown in the bottom left corner.} 
\label{dust}
\end{figure}


\subsection{Disk mass}

In Fig.\,\ref{dust}, we overlap a map of the observed dust continuum emission at 1.37\,mm (dashed red contours) with that of the modeled dust 
emission from the disk at the same wavelength (white contours). We draw the contours of the observed continuum emission down to a 20\% level
of the peak of 14.6\,mJy\,beam$^{-1}$. Below this level, corresponding to a radius of about 2000\,au from the star, the dust continuum emission 
suffers contamination from the outer envelope, which is outlined by the lower, dashed, white contour in Fig.\,\ref{mom} (corresponding to about the 10\%
peak level). Our model predicts a dust continuum peak of 12.4\,mJy\,beam$^{-1}$, in excellent agreement with the observed intensity.

We compare the disk mass estimated from the dust continuum emission with that predicted by the model. Following \citet{Hildebrand1983},
we estimate the disk mass from the continuum flux of 62.9\,mJy within the 20\% contour. We use an average gas temperature of 279\,K, determined
from the model between radii of 10\,au and 2000\,au, and assume a standard gas-to-dust mass ratio of 100 and a dust opacity of
1.0\,cm$^2$\,g$^{-1}$ at 1.3\,mm \citep[same for the model]{Ossenkopf1994}. From these parameters, we calculate a disk mass of 1.6\,M$_{\odot}$
and assign an uncertainty of $\pm$\,0.3\,M$_{\odot}$, corresponding to an uncertainty of $\pm$\,20\% of the measured flux. Our model predicts
a disk mass of 2.5\,M$_{\odot}$ within a radius of 3000\,au, which decreases to 1.4\,M$_{\odot}$ at 2000\,au. Although the model does not account for
a dusty envelope surrounding the disk, which might contribute to the observed dust flux, these values are in excellent agreement with the measured disk
mass. We conclude that very likely the disk mass is much less ($\sim$\,10\%) than the mass of the star.

\subsection{Comparison with \cite{Johnston2015}}

Previous ALMA observations by \citet{Johnston2015} reported about a circumstellar disk around a young star in AFGL\,4176, which has a mass
comparable to G023.01$-$00.41. These two sources have similar distances from the Sun and were observed with comparable resolution as well,
allowing for a direct comparison. They might represent two snapshots for the formation of an O-type star, with  G023.01$-$00.41 been younger,
and having the potential to become three-times more massive than AFGL\,4176, eventually. In the following, we highlight common features and
differences:

\begin{itemize}[noitemsep]

\item Both disks extend up to radii of $\sim$\,2000\,au from the central star, and have similar average temperatures of $\sim$\,200\,K, as determined
by LTE analysis of the CH$_3$CN K-ladders.  

\item There is a large difference ($5\times$) in the estimate of the disk mass for these two objects, of 1.6\,M$_{\odot}$ and 8\,M$_{\odot}$ for
G023.01$-$00.41 and  AFGL\,4176, respectively. However, this difference is due to the different dust opacities at 1\,mm assumed in the calculations,
of 1.0\,cm$^2$\,g$^{-1}$ and 0.24\,cm$^2$\,g$^{-1}$ respectively, whereas the integrated fluxes at the same wavelength are similar (50--60\,mJy).
We argue that the lower dust opacity used by \citeauthor{Johnston2015} is more appropriate for diffuse clouds \citep{Draine2003}. If consistent dust 
opacities are used, the two disks also have similar masses of $\sim$\,2\,M$_{\odot}$, which amount to $\sim$\,10\% of the central star mass. 

\item The main difference between the two disks is related to the disk kinematics. On the one hand, \citeauthor{Johnston2015} found Keplerian-like 
rotation up to the outer disk radii, meaning that nearly centrifugal equilibrium slows down any inward flow of mass through the disk. On the other hand, we 
find that the disk surrounding G023.01$-$00.41 is infalling and sub-Keplerian, and only at radii of a few 100\,au might approach centrifugal equilibrium.  
According to \citet[their Fig.\,4]{Kuiper2011}, larger Keplerian disks are expected at later times, whereas Keplerian rotation around the youngest stars
is confined to the inner disk regions. This evidence suggests that G023.01$-$00.41 is still in an active phase of accretion and much younger than AFGL\,4176.  

\end{itemize}

\section{Conclusions}\label{concl}

We report about Atacama Large Millimeter/submillimeter Array (ALMA) observations, at wavelengths near 1\,mm, of dense molecular gas and
dust in the vicinity of an O-type young star, with a linear resolution as good as 700\,au and a line sensitivity of 1\,K. We targeted the luminous
hot-molecualar-core G023.01$-$00.41 after selecting a best candidate disk-jet system from our previous observations.

We have resolved a (molecular) disk-jet system around a young star which currently attains a mass of 20\,M$_{\odot}$. We present a kinematic analysis
of the position-velocity diagrams of dense gas along the disk midplane, and compare them with the position-velocity diagrams simulated with a radiation
transfer model. We show that the disk is falling in close to free-fall and slowly rotating with sub-Keplerian velocities, from radii of about 2000\,au and down
to 500\,au from the central star, where we measure a mass infall rate of $6\times10^{-4}$\,M$_{\odot}$\,yr$^{-1}$. The disk mass is a small fraction
of the star mass ($\sim$\,10\%). Furthermore, we are able to image the jet emission which arises from the inner disk radii ($<$\,800\,au), and show that
its blueshifted emission leaves a strong imprint in the position-velocity diagrams.






\begin{acknowledgements}

This study  makes use of the following ALMA data: ADS/JAO.ALMA$\#$2015.1.00615.S. ALMA
is a partnership of ESO (representing its member states), NSF (USA) and NINS (Japan), together with NRC (Canada), NSC and ASIAA (Taiwan),
and KASI (Republic of Korea), in cooperation with the Republic of Chile. The Joint ALMA Observatory is operated by ESO, AUI/NRAO and
NAOJ. A.S. gratefully acknowledge financial support by the Deutsche Forschungsgemeinschaft (DFG) Priority Program 1573. RK and AK
acknowledge funding from the Emmy Noether research group on ``Accretion Flows and Feedback in Realistic Models of Massive Star Formation''
funded by the German Research Foundation under grant no. KU 2849/3-1.


\end{acknowledgements}


\bibliographystyle{aa}
\bibliography{asanna0518}



\begin{appendix}

\section{Disk model}\label{model}

We built up a semi-analytic model for a circumstellar disk surrounding a young star of 20\,M$_{\odot}$, in order to derive the
velocity profile of molecular gas flowing through the disk, and compare the expected \emph{pv}-diagrams with those obtained 
with the ALMA observations. Since the outflow axis is inclined by less than $30^{\circ}$ with respect to the sky plane, in our 
model we account for an inclination of $10^{\circ}$ between the disk plane and the line-of-sight, as a compromise. We explicitly 
note that the model does not change significantly within this range of inclinations.

Our semi-analytic models are discretized in order to be usable for the radiative transfer code RADMC-3D \citep{Dullemond2012}.
For this purpose, we use a regular two-dimensional, equatorial- and axial-symmetric spherical grid, with 152 cells in the radial
direction and 20 cells the in polar direction. The radial coordinates are logarithmically scaled such that the center regions are
well resolved. The grid ranges from the stellar radius to a radius of 0.05\,pc; the total image size is 0.1\,pc. Here, we consider
the radiation from the protostellar source as well as the emission from dust continuum and lines for a given molecule. We
further assume local thermodynamic equilibrium and prescribe a temperature distribution which is the same for the dust and
gas components. We configured RADMC-3D to use ray-tracing, ignoring dust scattering. The spectrum has a total width of
28\,km\,s$^{-1}$, divided into 80 channels centered around the corresponding line. From these images, we produce FITS-files
(Flexible Image Transport System) that are then readable by numerous astronomical software, including CASA.

We consider a constant disk aspect-ratio ($H/R$) of 0.1, and fix the disk density ($\rho$), temperature (T), and molecular
abundance ($\chi$) at a reference radius, R$_0$. The dependence of $\rho$, T, and $\chi$ on the disk radius is assumed to
be a simple power-law of exponents \emph{a}, \emph{t}, and \emph{q}, respectively. The disk density and temperature are
assumed to be, respectively, $1.0\times10^{-15}$\,g\,cm$^{-3}$ and 300\,K at a radius R$_0=500$\,au, following
\citet{Kuiper2013}. Their respective exponents, \emph{a} and \emph{t}, are $-1.5$ and $-0.4$, following
\citet{Gerner2014}. Note that, the choice of temperature is consistent with the (average) rotational temperature
(195\,K) of the CH$_3$CN\,($12_{\rm K}$--$11_{\rm K}$) spectra, determined within a radius of 3000\,au from the
HMC center by LTE analysis \citep{Sanna2014}. The dust properties are assumed constant across the disk, with a gas-to-dust
mass ratio of 100 and a dust opacity of 1.0\,cm$^2$\,g$^{-1}$ at 1.3\,mm \citep{Ossenkopf1994}. We set a CH$_3$CN
abundance relative to H$_2$ of $2\times10^{-8}$ at 500\,au, and assume a slow decrease with the radius similar to the 
temperature slope \citep[e.g.,][]{Sanna2014,Hernandez2014,Viti2004}.

The output files from the model were processed in CASA with the tasks ``simobserve'' and ``simanalyze'', in order to produce synthetic
maps with the same ALMA configuration (alma.cycle3.6.cfg) and observational conditions of the actual maps. In Figs.\,\ref{pvmod} and~\ref{dust}, we
present the results of this procedure for the CH$_3$CN\,($12_{4}$--$11_{4}$) line transition, at a frequency of 220.67929\,GHz,
and for the dust emission at the same frequency as the ALMA dust continuum map.

We note that the inner hole in the model of Fig.\,\ref{pvmod} (left) is due to the fact that we did not want to make any
assumption about the velocity profile in the outer regions of the disk. Most likely, in these outer regions the gas velocity will approach
the systemic velocity, filling the apparent hole in the model.

The main goal of our disk model is to show the range of velocities covered by different (ideal) velocity fields in the \emph{pv}-diagrams,
for which we assumed simple monotonic functions of $\rho$, T, and $\chi$, and constant dust properties as well. A posteriori, we can
comment on the accuracy of these assumptions based on the brightness distribution of the \emph{pv}-diagrams. On the one hand,
the modeled peak brightness of both the CH$_3$CN line and the dust continuum coincides with the observed values. This agreement
supports the choices for $\rho$ and T. On the other hand, in the modeled \emph{pv}-diagrams of Fig.\,\ref{pvmod}, the CH$_3$CN emission 
brightens the most at high velocities, which generally correspond to the inner disk regions. This is at variance with the observed
\emph{pv}-diagram, where the peak brightness is confined to the low velocities. This difference might be interpreted as the combination
of two effects: the dust properties change inside the disk, providing different line opacities at different distances from the star; inside the 
disk, the spatial distribution of the CH$_3$CN abundance does not follow a simple power-law of the disk radius.   

\end{appendix}

\end{document}